\documentclass{elsart}

\usepackage{harvard}

\usepackage{graphicx}

\usepackage{amssymb}

\def\url#1{{\ttfamily\def\/{/\discretionary{}{}{}}#1}}

\begin{document}

\begin{frontmatter}
\title{Stormy Weather and Cluster Radio Galaxies}


\author[Burns]{J.~O. Burns\thanksref{jb}}, 
\author[Burns]{C. Loken},
\author[Burns]{K. Roettiger},
\author[Burns]{E. Rizza},
\author[Bryan]{G. Bryan},
\author[Norman]{M.~L. Norman},
\author[Gomez]{P. G\'omez},
\author[Owen]{F.~N. Owen}

\thanks[jb]{E-mail: burnsj@missouri.edu}

\address[Burns]{Department of Physics \& Astronomy and Office of Research, University of
Missouri, Columbia, MO 65211}
\address[Bryan]{Department of Physics, M.I.T., Cambridge, MA 02139}
\address[Norman]{Department of Astronomy, University of Illinois, Urbana, IL 61801}
\address[Gomez]{Department of Physics \& Astronomy, Rutgers University, Piscataway, NJ
08855}
\address[Owen]{National Radio Astronomy Observatory, P.~O. Box O, Socorro, NM 87801}

\begin{abstract}

New adaptive mesh refinement N-body + hydrodynamics numerical simulations are used to
illustrate the complex and changing cluster environments in which many radio galaxies live and
evolve. Groups and clusters of galaxies form at the intersections of filaments where they
continue
to accrete gas and dark matter to the present day.  The accretion process produces shocks,
turbulence, and transonic bulk flows forming a kind of {\it{stormy weather}} within the
intracluster medium (ICM).  Radio sources embedded within the stormy ICM form distorted,
complex morphologies as observed in recent VLA cluster surveys.  We show that the bending of
wide-angle tailed radio sources can be understood as the result of recent cluster-subcluster
mergers.  We use new MHD simulations to illustrate how cluster radio halos can be formed by
the shocks and turbulence produced during cluster mergers.  Finally, we discuss new
observations of distant Abell clusters that reveal a class of weak radio sources, probably
starbursts, likely produced during the formation of the clusters as they accrete material from the
supercluster environment.

\end{abstract}

\end{frontmatter}

\section{The Environments Within Which Radio Galaxies Live \& Evolve}
\label{environ}

The environments of radio galaxies change dramatically over time.  Galaxies form within a
large-scale cosmic web that contains $>$100 Mpc-length filaments, composed of dark and
baryonic
matter, and huge voids of comparable size.  Groups and clusters, where nearly all galaxies reside,
form at the intersections of filaments.  Clusters continue to evolve even at the present epoch by
accreting gas and galaxies from connecting filaments and by merging with other clusters/groups
(roughly every few Gyrs).

\begin{figure}
\begin{center}
\includegraphics*[width=10cm]{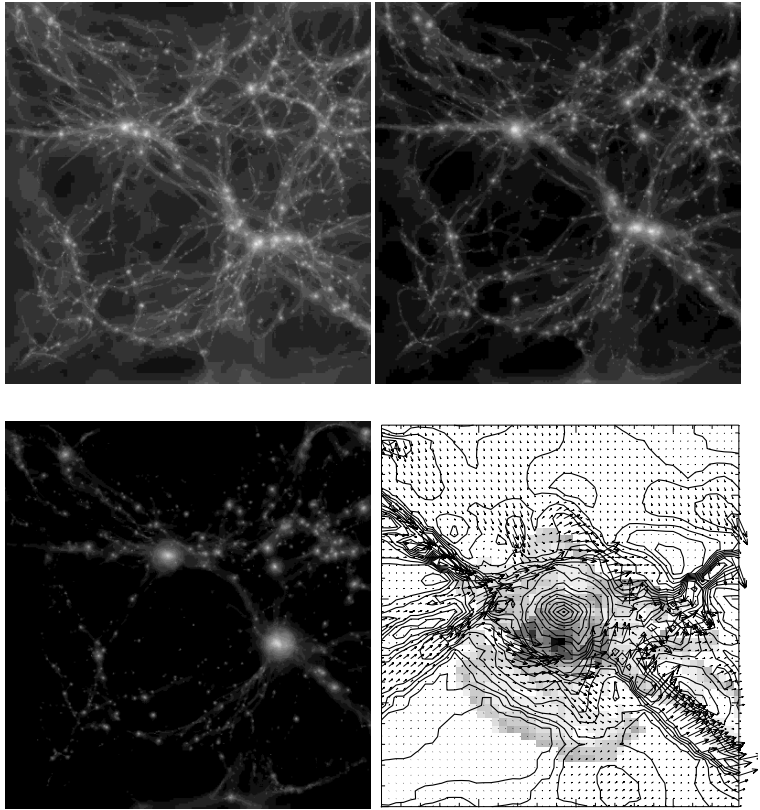}
\end{center}
\caption{AMR simulation of the evolution of rich clusters of galaxies in 
a $\Lambda$CDM universe.  The top two panels ($z=1$ and $z=0.5$) and bottom 
left panel ($z=0$) display the evolution of gas density within
a subvolume with sides of length $\sim50$h$^{-1}$ Mpc.
The bottom right panel shows gas density contours, momentum vectors
(below a minimum threshold), and grey-scale temperature (for gas
between 10$^6$ and 10$^8$K) in a 20h$^{-1}$ Mpc region centered
on the largest of the two $z=0$ clusters.
The entire computational volume has sides of 256 h$^{-1}$ Mpc.  
The mass resolution is $\sim$10$^{10}$ M$_{\odot}$ and the spatial 
resolution is $\sim$15.6 h$^{-1}$ kpc. }
\label{fig_amr}
\end{figure}

In Fig. \ref{fig_amr}, we illustrate the evolution of the gaseous environs of two Coma-like rich
clusters using a newly developed adaptive mesh refinement (AMR) numerical simulation
\cite{Loken99}.  This simulation is for a $\Lambda$CDM closed universe.  The AMR code uses
a series of grids that adaptively refine to higher resolution in proportion to the local density. 
Thus, high spatial dynamic range (8000:1) and excellent  resolution in the cluster core
($\approx$16 kpc) are achieved. 
The simulations reveal complex evolution within groups and clusters.  For example, the total
cluster mass and gas density doubles between $z=0.5$ and $z=0$ as a result of on-going mergers
and accretion from filaments.  The dark matter density profile maintains a general power-law
form {\citeaffixed{NFW97}{e.g.}} whereas the gas profile has a definite core (well-fit by a
$\beta$-model) produced from shock-heating and expansion of the central region.  Clusters have
substantial asymmetric temperature and pressure structure even at $z=0$; such predicted
temperature substructure will soon be tested with new observations from the 
{\it Chandra} and
{\it XMM} satellites.  These simulations, along with earlier numerical models
{\citeaffixed{Roett96}{e.g.}}, show that substantial bulk motions with velocities $\ge$1000
km/sec can be produced from the accretion and mergers in clusters (see Fig. 2).

These simulations suggest that the ICM within groups and clusters is {\it stormy}, violent, filled
with shocks, high winds, and turbulence.  The low density plasmas associated with extended
radio sources embedded within groups and clusters will act as windsocks and barometers of this
cluster weather \cite{Burns98}.

\section{Wide-Angle Tailed Radio Galaxies}
\label{WAT}

Wide-angle tails (WAT) are V-shaped, large diameter (0.2-1.2 ${h_{75}^{-1}} $Mpc) radio
sources most often associated with D/cD or giant ellipticals at the optical centers of clusters. 
They lie at the FR I/II interface.  Multifiber spectroscopy has confirmed that the
WAT galaxies are generally moving very slowly ($<$100 km/sec) relative the cluster velocity
centroid (\citeasnoun{Pink93}, \citeasnoun{Pink99}).  Such slow motion is insufficient to bend
the jets/tails of WATs to their observed curvature and, thus, an interesting puzzle emerges
concerning the origin of the shape of these relatively powerful cluster radio sources
\cite{Eilek84}.

Following the launch of {\it ROSAT}, we observed a statistical sample of WAT clusters using
the PSPC detector with $\approx$30" resolution \cite {Gomez97}.  These x-ray images
revealed statistically significant substructure in 90\% of the WAT clusters.  Furthermore, this
substructure generally consists of x-ray elongations that align closely with the directions of the
radio tails.

\begin{figure}

\includegraphics*[width=6cm]{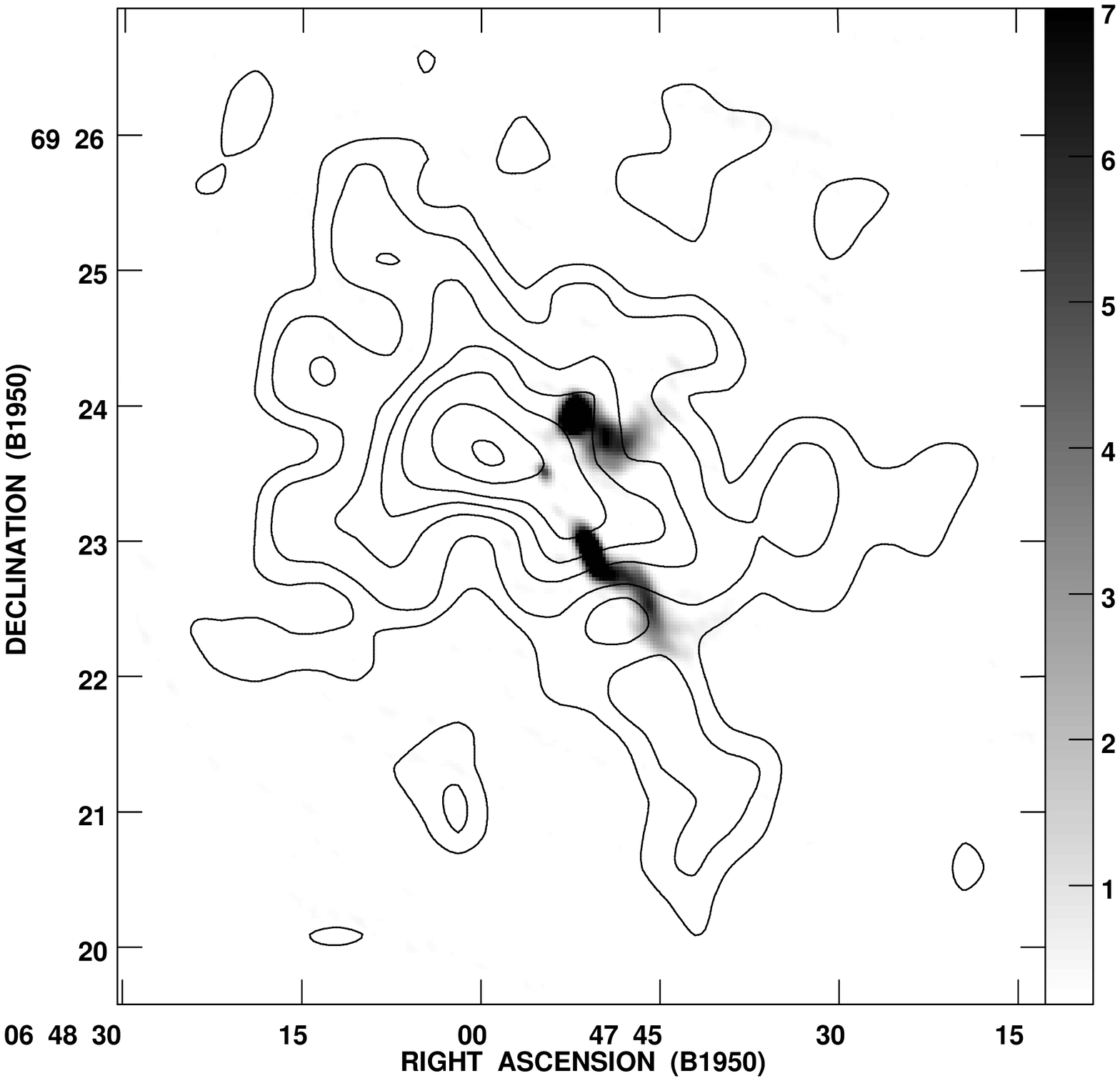}
\hspace{0.5in}
\includegraphics*[width=6cm]{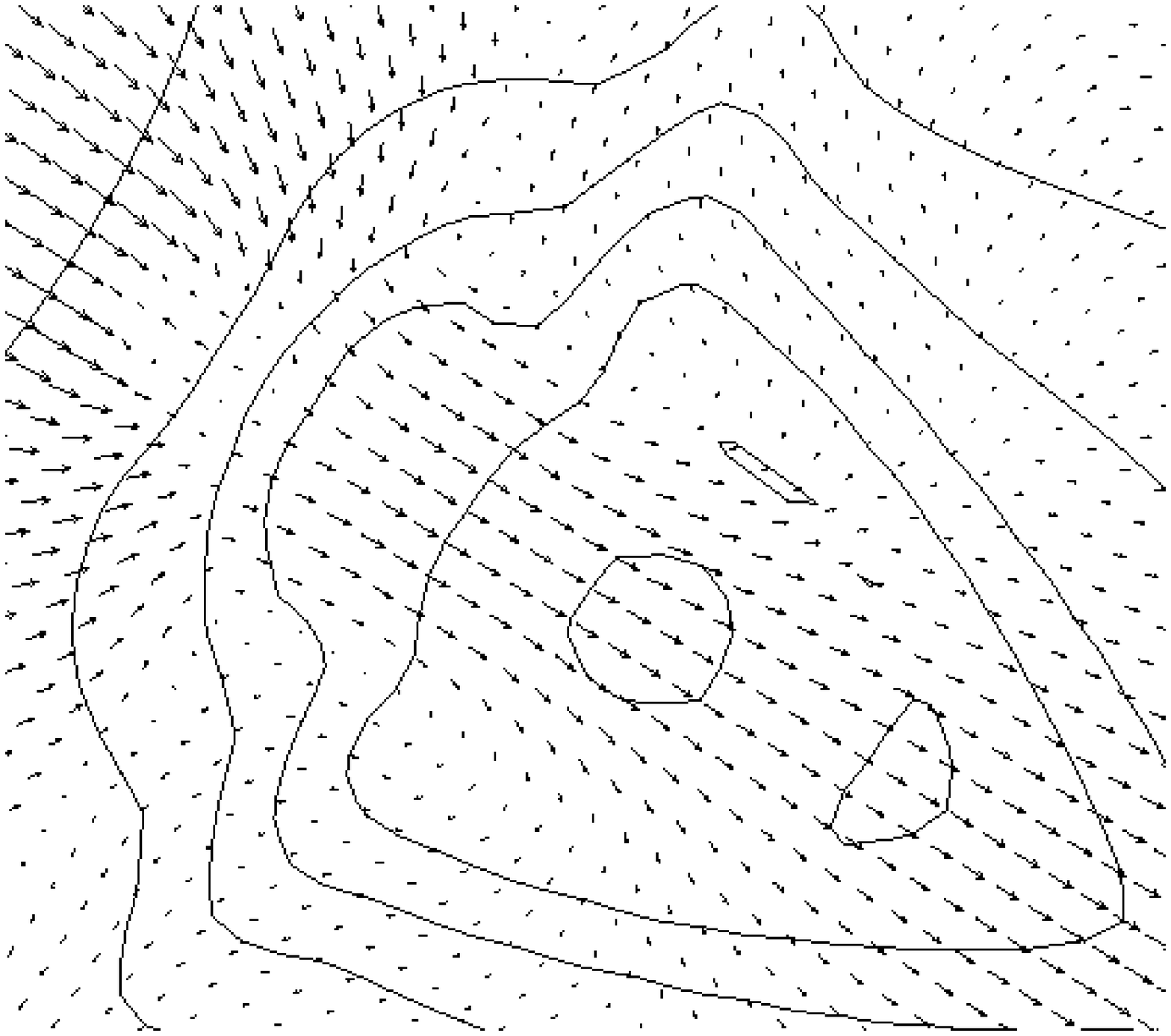}

\caption{Left: Overlay of 6 cm greyscale radio map onto a {\it{ROSAT}} PSPC image of A562. 
Right: Synthetic x-ray contours and velocity vectors from a cluster-subcluster simulation with a
4:1 mass ratio at 0.5 Gyrs after the subcluster passed through the cluster core.  Field is 2
$h_{75}^{-1}$ Mpc on a side with the longest vector of 1941 km/sec.  Note the bulk flow is
along the merger axis and coincides with the x-ray elongation.}
\label{fig_wat}
\end{figure}

In Fig. \ref{fig_wat}, we compare x-ray/radio observations of a WAT cluster with the results of
an N-body/hydro simulation of a cluster undergoing a merger with a smaller group of galaxies
\cite{Roett96}.  These simulations strongly suggest that the WAT tails are bent by motion of the
ICM past the radio galaxies arising from bulk flows following the merger between clusters. 
This contrasts with the traditional model of bending tailed radio sources via motion of the radio
galaxy through a static ICM.

This model has found additional support from the recent observation that there is a strong
alignment between the directions of WAT tails and the local supercluster axis as defined by the
distribution of nearby Abell clusters \cite{Nov99}.   Thus, we believe that the large scale
environment (i.e., accretion from filaments) is having an important direct impact on the central
regions, including radio sources, within clusters.

\section{Radio Halos and Cluster Mergers}
\label{halos}

Cluster radio halos are relatively rare (\citeasnoun{Giovan99} identify
29 candidate halo/relic souces), have steep radio spectra ($\alpha
\ge$1.0), and $\mu$G magnetic fields over very large volumes (halo diameters are often $\ge$1
Mpc).  These diffuse radio sources are not associated with any individual galaxy but are truly a
cluster phenomenon.  They are generally found in very rich (Abell richness class $\ge$2), X-ray
luminous clusters.  Some form of in-situ particle acceleration and magnetic field amplification is
required to maintain the synchrotron emission on such large scales.  Turbulent galaxy wakes 
have insufficient kinetic energy to power the radio halos {\citeaffixed{deY92}{e.g.}}.

Recently, it has been found that most radio halos appear to lie in clusters that are undergoing
cluster-cluster mergers (e.g., Coma, A2255, A2256, A3667) {\citeaffixed{Burns98}{e.g.}}. 
This
seems to
suggest that such mergers play an integral role in the formation and evolution of the radio halos. 
Cluster mergers do have sufficient kinetic power, comparable to the total cluster thermal energy
loss rate ($10^{46}$ ergs/sec), to energize halos with typical radio luminosities of
$\sim$10$^{42}$ ergs/sec.

One of us (Roettiger) has developed a new N-body/MHD code for modeling the evolution of
magnetic fields during cluster mergers \cite{RSB99}.  The ICM and B-fields are evolved using
the Eulerian, finite difference code ZEUS and the particles are evolved using a PM
(particle-mesh) algorithm.  A diffusive shock acceleration model, whereby the shock strength
determines
the power-law slope of  the injected particle energy spectrum, was used to produce simulations of
radio emission during cluster mergers.   This technique was applied to Abell 3667 and found to
reproduce the observed large scale radio ``arches" \cite{Rott97}, which we identify as the sites of
shocks formed during the merger between clusters \cite{RBS99}.  The numerical model also
reproduces the elongated cluster x-ray emission and is consistent with the observed bimodal
distribution of cluster galaxies in A3667.

\begin{figure}
\begin{center}
\includegraphics*[scale=0.7,angle=0]{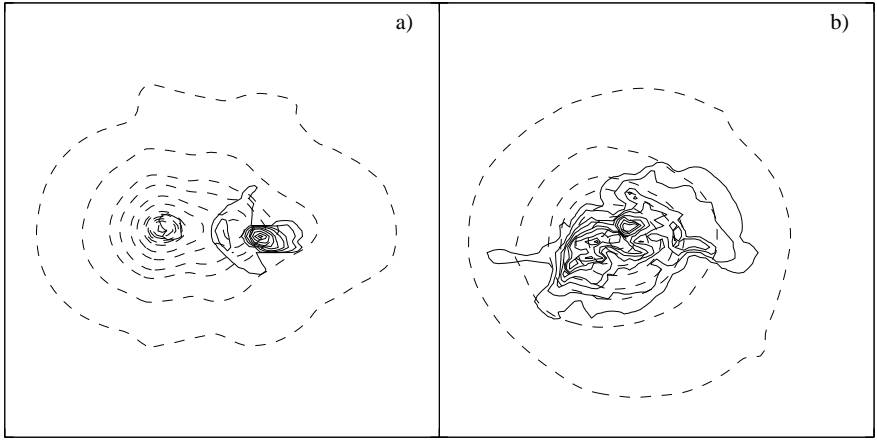}
\end{center}
\caption{Preliminary results from a numerical simulation of 
radio halo evolution, a) 0.4 Gyr before core passage 
and b) 6 Gyrs after core passage. Dashed contours represent the
soft x-ray emitting thermal ICM while the solid contours represent the sychrotron
emission from the relativistic plasma. Each panel is $\sim$6 Mpc on
a side.}
\label{fig_halo}
\end{figure}

We have now extended these numerical simulations to include more sophisticated
diffusive shock and turbulent particle acceleration along with synchrotron aging following the
approach of \citeasnoun{JJ99}.  This technique treats the relativistic electrons as multiple fluid
components each representing a range in particle momenta.  In Fig. \ref{fig_halo}, we show
preliminary results of the evolution of the radio
and x-ray emission during a cluster-cluster merger.  Lending support
to the evolutionary scenario suggested
in \citeasnoun{RSB99}, these new simulations show the radio emission in early epochs to
be dominated by shock acceleration producing edge-brightened, off-center relics 
resembling those
in A2256 and A3667.  At later times, the shocks diminish in strength as the cluster relaxes and
turbulence drives the particle acceleration producing center-filled halos that resemble Coma. 

\section{A Possible Radio Analog of the Butcher-Oemler Effect}
\label{BO}

A final possible manifestation of accretion and mergers in the cluster environment on radio
sources is illustrated using new observations of weak radio galaxy emission in more distant rich
clusters.  We used the VLA to deeply image $z$$>$0.2 very rich clusters (Abell richness class
$\ge$2) down to 20-cm radio powers of $\approx$10$^{21.8}$ W/Hz.  At such low powers, the
radio luminosity function (RLF) is dominated by star formation and weak AGNs.  Each of these
clusters was also detected by the {\it ROSAT} all-sky survey at 0.5-2.0 keV energies.

At 20-cm powers $\ge$10$^{23}$ W/Hz, we find that the RLF and the spatial distribution of
radio sources look virtually identical between samples of nearby ($z \le 0.09$) and more distant
($z=0.2-0.4$) rich clusters.  These sources are principally traditional AGNs that are associated
with galaxies generally concentrated within the cores of clusters. \citeasnoun{Stocke99}
confirms this lack of change in the characteristics of powerful radio sources in even more distant
clusters selected from the {\it Einstein} medium sensitivity x-ray survey.  Thus, in spite of the
dramatic evolution of the cluster environment, the distribution and luminosities of powerful radio
sources remain relatively unaffected by these events.

On the other hand, there appear to be some very interesting, significant changes in the population
of low power radio sources over these same redshift intervals.  For radio sources with
$P_{20cm} \le 10^{23}$ W/Hz, the spatial distribution of weak radio galaxies is much broader
extending over several Mpcs rather than being confined to the cluster core (several hundred kpc) 
{\citeaffixed{Morrison99,Rizza2000}{e.g.}}.  This extent is more on the supercluster than the
cluster scale. 
Furthermore, in the case of some individual clusters such as A2125, the number of weak radio
sources is much greater (factor of 5) than that in comparable richness nearby clusters
\cite{Owen99}.  There is also some suggestion that this trend of increased numbers of weak
radio sources becomes stronger at higher redshifts ($z\sim 0.4$)  \cite{Morrison99}.

These low power radio sources are likely to be starburst galaxies.  Such starburst emission may
be triggered as groups of galaxies accrete onto richer clusters.  The x-ray emission observed by
{\it ROSAT} for many of these clusters is clumpy and suggests that on-going mergers are
common at these epochs \cite{Rizza99}, as suggested by the numerical simulations.  Thus, we
may be seeing a radio analog of the Butcher-Oemler effect in these rich clusters.

\section{Conclusions}
\label{conclude}
          
The gaseous environment within groups and clusters of galaxies is filled with ``stormy weather''
including shocks, high winds, and turbulence.  This violent weather is produced by mergers
between clusters and groups, and by on-going accretion of matter from supercluster filaments
that connect clusters to the large-scale structure of the universe.  New AMR numerical
simulations reveal great complexity within the ICM that has important consequences for the
extended radio sources that lie within clusters.

Wide-angle tailed radio sources are likely bent into a V-shape as the result of bulk flows of gas
moving by the central cluster radio galaxies at velocities of $\ge$1000 km/sec.  This dynamic
pressure arises from mergers and accretion of gas from large-scale filaments.  The observed
correlations between x-ray elongations of the ICM with the WAT radio tail directions, and
between the supercluster axis and the tail directions support this hypothesis.

New N-body/MHD simulations, which include diffusive shock and turbulent acceleration of
relativistic electrons, are able to reproduce the morphologies of radio halos resulting from
mergers between clusters.  A possible evolutionary scenario is suggested whereby early-stage
mergers produce halos whose morphologies are dominated by the effects of shocks (e.g., A3667)
whereas late-stage mergers produce center-filled halos dominated by turbulence (e.g., Coma).

Finally, we have observed strong evolution in a population of weak radio sources in distant rich
clusters.  These weak sources are likely to be starburst galaxies.  The x-ray emission from these
clusters is often asymmetrical and clumpy.  This suggests that the starbursts may be stimulated
by
galaxies in groups infalling from supercluster filaments onto rich clusters, similar to what has
been proposed to explain the excess of blue galaxies in more distant clusters (i.e., Butcher-
Oemler effect).

We thank Michael Ledlow, Anatoly Klypin, and Wolfgang Voges for their collaboration on
aspects of this research.  This work was supported by the NSF (AST-9896039) and by NASA.


\end{document}